\documentclass[preprint2]{aastex}

\usepackage{apjfonts}

\usepackage{graphics}
\usepackage{psfig,natbib}
\usepackage{psbox}

\newcommand{\kms}	{km~s$^{-1}$}
\newcommand{\h}   	{$h^{-1}\,$~kpc}
\newcommand{\mpc}       {$h^{-1}\,$~Mpc}
\newcommand{\etal} 	{{et~al.}}

\newcommand{\lya}	{Ly$\alpha$}

\newcommand{\push}[1]   {\multicolumn{1}{c}{#1}}
\lefthead{Absorbers towards Q2302+029}
\righthead{D. V. Bowen et al.}

\received{14-Jun-2000}
\accepted{11-Sep-2000}

\begin{document}

\title{Where are the absorbers towards Q2302+029?}

\author{David V.~Bowen$^{1,2,3}$, 
Raul Jimenez$^{4,5}$,  Edward B.~Jenkins$^{1}$
and Max Pettini$^6$}

\affil{ $\:$}

\affil{$^1$ Princeton University Observatory, Princeton, NJ 08544} 

\affil{$^2$ Royal Observatory, Blackford Hill, Edinburgh EH9 3HJ, UK}

\affil{$^3$ Department of Physics \& Astronomy, Johns Hopkins
University, Baltimore, MD 21218}

\affil{$^4$ Institute for Astronomy, University of Edinburgh,
Blackford Hill, Edinburgh EH9 3HJ, UK}

\affil{$^5$ Current address: Rutgers University, Dept.~of Phys.~\&
Astron., 136 Frelinghuysen Rd., Piscataway, NJ 08854}

\affil{$^6$Institute of Astronomy, Madingley Rd.,
Cambridge CB3 0HA, UK}

\begin{abstract}

We present images and spectroscopy of objects close to the sightline
of Q2302+029 in order to search for galaxies responsible for the
remarkable $z\:=\:0.7$ high-ionization absorption line system found by
\citet{Janu96}.  This system shows `normal' narrow O~VI, N~V, and C~IV
lines superimposed on broader (3,000$-$5,000~\kms\ wide), unsaturated
absorption troughs some 56,000~\kms\ away from the QSO emission
redshift ($z_{\rm{em}}\:=\:1.052$).  Despite reaching sensitivities
sufficient to detect $1/10 L^*$ galaxies in the optical and $1/20 L^*$
in the infra-red, we are unable to detect any obvious bright galaxies
which might be responsible for the absorption beyond $\approx 6$~\h\
of the sightline. This supports the hypotheses that the absorption is
either intrinsic to the QSO, or arises in intracluster gas. Adopting
either explanation is problematic: in the first case, `associated' absorption
at such high ejection velocities is hard
to understand, and challenges the conventional discrimination
between intrinsic and intervening absorbers; in the second case, the gas
must reside in a $\sim\:40$~\mpc\ long filament aligned along the line of
sight in order to reproduce the broad absorption. Since the absorption
system is unusual, such a chance alignment might not be unreasonable.

Spectroscopy of objects beyond the immediate vicinity of the QSO
sightline reveals a galaxy cluster at $z\:=\:0.59$, which coincides
with strong \lya\ and more narrow high ionization lines in the quasar
spectrum. Here too, the lack of galaxies at distances
comparable to those found for, e.g., \lya -absorbing galaxies,
suggests that the absorption may arise from intracluster gas unassociated
with any individual galaxies.

\end{abstract}

\keywords{quasars:absorption lines---quasars:individual 
(Q2302+029)---large-scale structure of universe}

                      \section{Introduction}

The transition of the O~VI $\lambda\lambda 1032,1037$ absorption
doublet seen in the spectra of QSOs has long been of interest because
of the difficulty in producing O~VI ions by photoionization alone. The
ionization of O~V to O~VI requires photons with energies above 114~eV,
which are rare except in the radiation field of QSOs and AGN.
Traditionally, it has been thought that O~VI absorption probably marks
the existence of collisionally ionized gas, since under equilibrium
conditions the fraction of oxygen residing as O~VI in a plasma peaks
at a temperature of $T\:\sim\:3\times10^{5}$~K \citep{SuthDopi}. More
recently, with the explosive development of hydrodynamical simulations
which follow the physical conditions of gas as galactic large-scale
structure forms, it appears that a substantial fraction of baryons may
be in a shock heated phase at temperatures of $10^5-10^7$~K as a
result of gravitational collapse \citep[see, in particular,][and
refs.~therein]{CenOstr}. O~VI absorption lines have been hard to
detect at high redshifts due to confusion with the \lya -forest
\citep{LuSava} but access to ultraviolet wavelength regions with the
{\it Hubble Space Telescope} ({\it{HST}}) has made it possible to
detect such systems at redshifts $z\:<\:1$ as the forest thins out
\citep{BurlTytl, KP6, KP13}.  Moreover, recent work suggests that O~VI
systems are remarkably common at low redshift, and that the lower
limit to the cosmological mass density contributed by such systems
today may be comparable to, or exceed, that of stars and cool gas in
galaxies, as well as X-ray emitting gas in galaxy clusters
\citep{trip00a,trip00b}.

In trying to understand the origin of the many types of QSO absorption
line systems, the most obvious question to ask is whether the
absorption is directly associated with galaxies. To date, the answer
to this question has been surprisingly ambiguous for a wide range of
absorber types. Despite initial detections of Mg~II absorbing galaxies
which seemed to firmly establish the existence of $\sim\:40$~\h
\footnote{$h\:=\:H_0/100$, where $H_0$ is the Hubble constant, and
$q_0\:=\:0.5$ is assumed throughout this paper} radii halos for bright
galaxies \citep{jb1, BerjBois, Stei94}, it remains unclear whether
individual galaxies are responsible for \lya\ lines
\citep{LBTW,Chen98,Morr93,Stoc95,BBP96,LeBr96,Shul96,vanG96,BPB98,Trip98},
while the diversity of galaxy types associated with damped \lya\
systems has been a surprise \citep{LeBr97,Lanz97,RaTu98}.
Establishing whether O~VI absorption is linked to galaxies is
important for establishing whether absorption is indeed due to violent
processes within the ISM of individual galaxies or the consequence of
the growth of large-scale structure in the universe.  The effort to
understand these systems is now even more timely with the launch of
the {\it Far Ultraviolet Spectroscopic Explorer} ({\it{FUSE}}) which
will characterize the distribution of O~VI absorbing gas in the Milky
Way disk and halo, and in the Magellanic Clouds, since the data
obtained will provide an important baseline for understanding the
higher redshift systems.

\begin{figure}[th]
\hspace*{-1.1cm}
\psbox[xsize=0.55\textwidth]{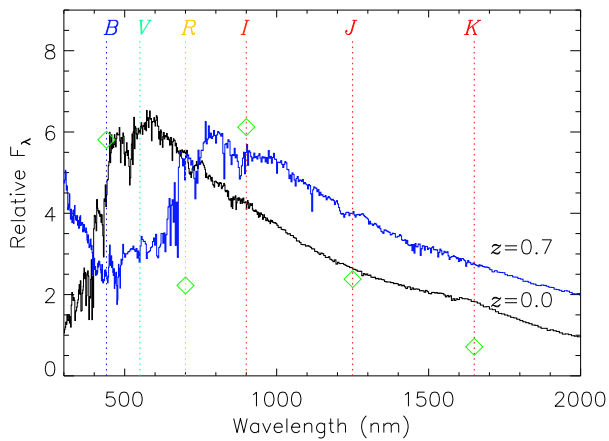}
\caption{\label{fig_sed1} The dark curve shows the SED of a  10~Gyr
old---i.e. present day---galaxy with metallicty of $Z\:=\:0.2\:Z_\odot$ 
at $z\:=\:0$. The lighter curve shows the same
galaxy, when only 5~Gyr old at $z\:=\:0.7$. The absolute
normalization of the SEDs are arbitrary. The peak of the
redshifted galaxy has shifted from between the $B$- and $V$-bands to
between the $R$- and $I$-bands; with both optical and IR magnitudes
measured for such a galaxy, it is possible to constrain the wavelength of
the SED peak, and hence the object's redshift. The diamonds
denote the {\it relative} flux limits of our data towards
Q2302+029. They are scaled such that $F_{\lambda}$($I$) lies at the
peak of the galaxy's SED, and are discussed in the text.}
\end{figure}

For these reasons, we have begun a survey to search for the absorbers
responsible for $z\:<\:1$ O~VI absorption systems. Our approach is
similar to those of previous studies --- deep imaging of the field
around the QSO followed by spectroscopic confirmation of detected
objects. However, many of the O~VI systems detected by {\it{HST}} are
at redshifts almost inaccessible to spectroscopic observations with
4~m class telescopes.  We have therefore aimed to obtain multicolor
information --- in particular, deep infra-red (IR) images --- in order
to estimate galaxy redshifts photometrically. For example, in
Figure~\ref{fig_sed1} we show the spectral energy distribution (SED)
of a late-type galaxy at $z\:=\: 0.7$. (We discuss these models in
more detail in \S 3.3). The figure shows that the peak of a galaxy's
SED at these redshifts occurs at wavelengths between the optical and
the IR; combining photometry in these two regimes should enable us to
constrain a galaxy's redshift.

In this paper we discuss some initial results of our survey, a search
for galaxies towards the radio-quiet QSO Q2302+029
($z_{\rm{em}}\:=\:1.052$). {\it HST} spectra obtained by
\citet{Janu96} show remarkable broad (FWHM$\:\sim\:3,000-5,000$~\kms )
O~VI, N~V, and C~IV absorption at $z\:=\:0.695$, with more typical
narrow-lines (including \lya , which is not seen as a broad component)
superposed at $z\:=\:0.7106$. Further, the system shows no
low-ionization species such as Si~II and Mg~II.  Jannuzi~\etal\
considered three possible causes for the absorption: gas ejected by
the QSO at a velocity of $\sim\:56,000$~\kms; gas from galaxies and/or
the intracluster medium of a cluster or supercluster; or gas from a
supernova remnant in a single galaxy at a redshift similar to that of
the narrow line system. Jannuzi~\etal\ were unable to distinguish
between these three possibilities with the available data; clearly, a
search for absorbing galaxies close to the line of sight is an
important first step in attempting to unravel the origins of this
unique system.

                \section{Observations and Catalogs}

                  \subsection{Imaging observations}

$J$- and $K$-band images of the field around Q2302+029 were taken
1998-Sept-11 \& 12 at the 3.8~m United Kingdom Infrared Telescope
(UKIRT) on Hawaii using the IRCAM3 camera. The detector was a 256x256
Indium Antimonide (InSb) array, each pixel 0.286 arcsec square, giving
a field of view of 73x73 arcsecs. Data were taken in NDSTARE mode,
such that a single dark exposure was followed by nine minute exposures
broken into nine 20x3 sec exposures in J and nine 10x6 sec
exposures in K. (The pixels of the InSb array are quickly filled by
photons because of the high sky background, so frequent read-out of
the array is required.)  After each one-minute exposure, the telescope
was shifted by 8 arcsec, forming a grid of nine pointings in one
nine-minute group, to enable production of an accurate flat-field (FF)
during the reduction of the data. When the nine-minute group
observation was completed, the center of the grid was shifted by 2
arcsec, and the nine-minute pattern repeated.  Total exposure times of
72 and 170 mins were obtained for the $J$- and $K$-band images
respectively. Observations of `faint' UKIRT standard stars were also
taken shortly before those of the QSO field.

\begin{table*}[ht]
\begin{center}
\begin{tabular}{ccccccccc}
\multicolumn{7}{c}{Table 2. Limiting magnitudes in the field of Q2302+029 \label{tab_lims}}\\
\tableline\tableline
    & $m_{\rm{obs}}$     & $M_{\rm{obs}}$$^a$   & $k+EC$-corr$^b$  
& $M$$^c$  & $M^*$$^d$ & $L/L^*$ \\
\tableline
$B$ & 25.2 & $-16.8$   &  1.0   & $-17.8$  & $-19.3$   & 0.3     \\
$R$ & 24.5 & $-17.4$   &  0.6   & $-18.0$  & $-20.2$   & 0.1     \\
$I$ & 22.8 & $-19.1$   &  0.6   & $-19.7$  & $-21.8$   & 0.1     \\  
$J$ & 22.6 & $-19.3$   &  0.4   & $-19.7$  & $-22.7$   & 0.06     \\
$K$ & 21.1 & $-20.8$   &  0.2   & $-21.0$  & $-23.6$   & 0.09     \\
\tableline 
\multicolumn{7}{l}{$^a$ observed absolute magnitudes assuming $z=0.7$, $q_0=0.5$;}\\
\multicolumn{7}{l}{all values of $M$ above assume $h\:=\:1$}\\
\multicolumn{7}{l}{$^b$ $k$+EC-corrections for $z\:=\:0.7$---see text}\\
\multicolumn{7}{l}{$^c$ $k$+EC-corrected absolute magnitudes}\\
\multicolumn{7}{l}{$^d$ $M^*$ from: 
$B$---\citet{Marz98}; 
$I$---\citet{Metc98};}\\
\multicolumn{7}{l}{  $\:R$---\citet{Lin96}; $J$---see text; $K$---\citet{Love00}}
\end{tabular}
\end{center}
\end{table*}

To reduce the data, we used IRAF\footnote{The Image Reduction and
Analysis Facility (IRAF) software is provided by the National Optical
Astronomy Observatories (NOAO), which is operated by the Association
of Universities for Research in Astronomy, Inc., under contract to the
National Science Foundation.} to first subtract the dark frame from
each of the nine-minute sub-exposures. Because the sky varies over
short time intervals, the FF used to flatten the sub-exposures of a
nine-minute pointing needed to be constructed from only the one-minute
exposures comprising the group. Although the median of nine frames
with high background counts produces an adequate signal-to-noise, the
small number used means that `residuals' in the FF occur at the
positions of the brightest objects in the frame. To overcome this, we
defined circular masks at the positions of all bright objects in the
(initial) flat-fielded image, using the {\tt ximages} package of the
IRAF PROs package. A new, more accurate, FF was constructed by forming
the median of the nine individual frames, but with pixels in the
masked regions excluded.  Each
sub-exposure was then shifted to a common reference position and a
final coadded image produced. A small non-linearity correction was
made, and a photometric zero-point calculated using
the standard stars.

Harris $B$-, $R$-, and $I$-band images were taken with the Wide Field
Camera (WFC) at the 2.5~m Isaac Newton Telescope on La Palma on
1998-Nov-22. The detector consists of four thinned EEV 2Kx4K CCDs each
with a plate scale of 0.33 arcsec pixel$^{-1}$ and a coverage of
22.8x11.4 arcmin. Exposures of 400 sec were taken at pointings
differing by ten arcsecs. Total exposure times of 53.3, 26.7 and 20.0
mins were obtained for the $B$, $R$, and $I$-bands respectively, and
the data were bias-subtracted and flat-fielded with dome flats in the
standard manner. The $I$-band data suffered from severe fringing; to
remove this, the flat-fielded $I$ frames were medianed to produce a
fringe frame with a mean of zero which was scaled to, then subtracted
from, each $I$-band frame.  Photometric zero points were derived from
standard star exposures bracketing the science data, and a correction
was applied for non-linearities in the electronics.

       \subsection{Photometric catalogs and magnitude limits}

To detect and catalog galaxies in our images, we ran the {\tt
sextractor} routines \citep{BeAr96} on each of the five science
frames.  Catalogs of objects from each passband were merged, and
appropriate extinction corrections were applied to the derived
magnitudes.

In Table~1 we list all the objects detected within a radius of 0.67
arcmins; this corresponds to 160~\h\ at $z\:=0.7$, which is the radius
found by \citet{LBTW} for \lya -absorbing halos, and which we
therefore consider to be the radius of interest.  In principle, O~VI
absorbing halos might be larger than this, but the examination of the
objects closest to the line of sight naturally received the highest
attention. Objects without $J$ and $K$ entries in Table~1 were not
within the IRCAM field of view.

The magnitude limits for each passband are given in
Table~2. The limits are based on both a search for faint
objects in the data frames, as well as a $2\sigma$ limit calculated
from the signal-to-noise of the sky in a circular aperture with a
radius twice the seeing FWHM. These two estimates were found to agree
closely, despite the obvious problem that the magnitudes of the
faintest objects in a frame also have the largest errors.  Due to the
large area of the WFC CCDs, it was also possible to construct a
histogram of the distribution of galaxy magnitudes in the optical and
observe the magnitude at which the distribution begins to
decline. This is a partial indicator of a magnitude limit, simply
because the numbers of galaxies should continue to increase with
increasing magnitude.  Unfortunately, this estimate suffers from the
fact that galaxy type, inclination, etc., also play a role in where
the surface density of galaxies begins to decrease, so it does not
offer a definitive measure of limiting magnitude. However, it does
give an indication of the {\it completeness} in the image, and by
plotting such histograms of galaxy magnitudes, we deduce that we are
complete in $R$ and $B$ to about 0.5 mags brighter than the limits
given in Table~2. In $I$, the distribution is less well
peaked due to the smaller number of galaxies detected, and the
completeness limit is about the same as the magnitude limit measured
as described above.

                 \subsection{Spectroscopic observations}

Spectroscopic observations of objects in the field of Q2302+029 were
obtained during the last run before de-commissioning of the Low
Dispersion Survey Spectrograph-2 (LDSS-2) at the William Herschel Telescope
(WHT) on La Palma, 1999-Aug-7, 9, \& 10. A SITe1 2148x2148 CCD was
used in conjunction with the medium/red grating which gave a dispersion
of $\simeq\:5.0$ \AA\ pix$^{-1}$ and resolution of $\simeq\:12.0$~\AA\
FWHM.

We selected all objects with $R=15.0-23.0$ within the central 0.75
arcmins ($\equiv\:180$~\h\ at $z\:=\:0.7$). To this we added objects
with $R\:=\:20.0-22.0$ within an annulus of radius 0.75$-$4.0 arcmins.
Our aim was to focus on objects closest to the QSO, while using
the brighter sample to fill up slits in the outer parts of the field
of view. In practice, because of the large number of objects in the
catalog, and the limited amount of space on a mask for slits, the area
was well sampled over the LDSS-2 field of view without any bias in
location. Three LDSS-2 masks were employed, each covering a 5 arcmin
radius field of view, and containing 21$-$27 1.5 arcsec wide and 10
arcsec long slits. We note that these did not include the full number
of galaxies selected, and we are therefore incomplete to any magnitude
limit.  Total exposure times varied for each mask due to observing
constraints: 6x, 4x, and 3x 1800 sec were recorded for the three
masks. We were unable to observe any standard stars to
provide flux calibration for the galaxies, but one radial velocity
standard was observed to improve the accuracy of the redshift
measurements.

The data were again reduced using IRAF, with each individual galaxy
spectrum (and associated arc) `cut' from the 2D frame. Although arcs
were taken before each set of mask exposures, we noticed shifts in the
wavelengths of sky lines from one exposure to the next within each
exposure set. Hence an offset was applied to each spectrum to set the
O~I $\lambda 6300.3$ sky line to its geocentric value. Redshifts were
estimated manually, and refined by cross-correlation techniques when
possible using the IRAF {\tt fxcor} package.

\begin{figure*}[ht]
\vspace*{-2.5cm}\centerline{\psfig{figure=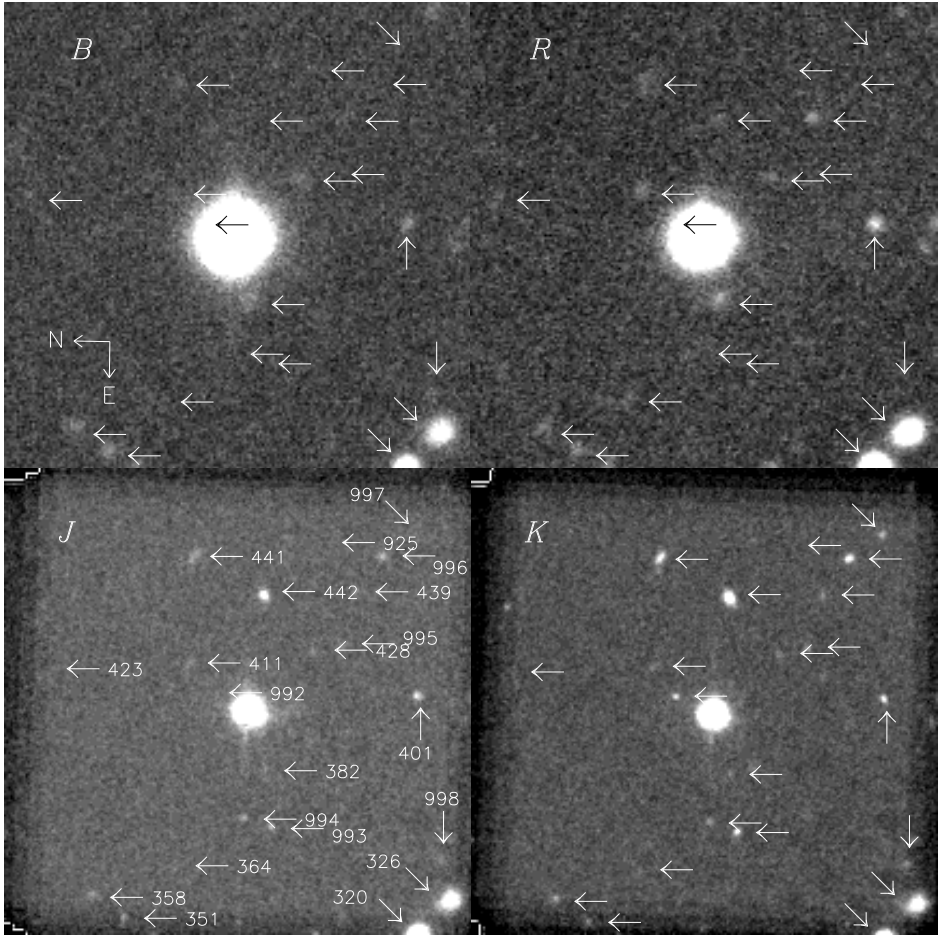,height=15.5cm,angle=0}}
\vspace*{-1cm}\caption{\label{fig_panel} 
Portions of the $B$-, $R$-, $J$- and $K$-band images around
Q2302+029, each 62x62 arcsec on a side (62 arcsec corresponds to 247~\h\ at $z\:=\:0.7$).
The positions of all detected
objects (even if detected in only a single band) are shown by arrows---for the
sake of clarity, their numerical designations are only plotted on the
$J$-band image. Although many detections are shown, most arise in
the $K$-band, and probably represent galaxies at $z>1$. Galaxy
401 is the only galaxy for which we have a spectroscopically confirmed
redshift, $z\:=\:0.365$.
\label{fig_panel}}
\end{figure*}

                         \section{Results}

\subsection{Imaging}

A montage of the inner 30 arcsecs around Q2302+029 in the $B$-, $R$-,
$J$-, and $K$-band is shown in Figure~\ref{fig_panel}. 
The most striking observation is that there are actually very few
galaxies within this radius. The lack of detections in the $J$-band is
the most telling, since these are our deepest images and are most
sensitive to the old stellar populations at $z\:=\:0.7$.

\begin{table*}[ht]
\begin{center}
\begin{tabular}{lccrccclrc}
\multicolumn{10}{c}{Table 3. Redshifts obtained in the field of Q2302+029}\\
\tableline\tableline
 &
\multicolumn{3}{c}{Position (J2000.0)} &
 &
 &
\push{$\theta$} &
 &
\push{$\rho$} &
  \\ \cline{2-3}
ID &
RA &
DEC &
\push{$B$} &
$R$ &
$I$ &
\push{($'$)} &
\push{$z$} &
\push{(\h )} &
$M_{\rm{obs}}(R)$ \\
\tableline
    65     & 23:04:56.77 &  03:13:42.8 &  23.1  & 22.0 & 21.5  &  3.529 & 0.589 & 800.6  &    $ -19.5$ \\
    91     & 23:04:55.15 &  03:10:55.5 &  22.0  & 20.7 & 20.4  &  2.678 & 0.3566& 487.0  &    $ -19.6$ \\
   117     & 23:04:54.70 &  03:12:59.5 &  22.7  & 21.6 & 21.5  &  2.719 & 0.590 & 617.2  &    $ -19.9$ \\
   135     & 23:04:54.06 &  03:13:40.4 &  22.5  & 20.7 & 20.0  &  2.962 & 0.3165& 504.0  &    $ -19.3$ \\
   164     & 23:04:52.79 &  03:13:34.0 &  22.1  & 21.2 & 20.8  &  2.653 & 0.426 & 527.5  &    $ -19.5$ \\
   251     & 23:04:49.49 &  03:12:05.2 &  22.7  & 21.9 & 21.0  &  1.172 & 0.7197& 282.2  &    $ -20.0$ \\
   265     & 23:04:48.55 &  03:11:25.5 &  23.1  & 20.2 & 19.0  &  0.958 & 0.5911& 217.6  &    $ -21.3$ \\
   291     & 23:04:48.32 &  03:12:42.7 &  22.8  & 21.5 & 21.2  &  1.259 & 0.5915& 286.0  &    $ -20.0$ \\
   325     & 23:04:47.21 &  03:09:25.8 &  24.2  & 21.8 & 20.8  &  2.406 & 0.327 & 417.1  &    $ -18.3$ \\
   354     & 23:04:46.21 &  03:10:35.4 &  23.4  & 20.9 & 19.6  &  1.220 & 0.5915& 277.1  &    $ -20.6$ \\
   359     & 23:04:46.39 &  03:15:08.6 &  24.0  & 21.2 & 20.2  &  3.392 & 0.6453& 792.4  &    $ -20.5$ \\
   365     & 23:04:45.93 &  03:12:56.9 &  22.0  & 21.3 & 20.8  &  1.202 & 0.058 &  55.1  &    $ -14.9$ \\
   368     & 23:04:45.67 &  03:12:24.4 &  23.5  & 21.4 & 20.6  &  0.660 & 0.59  & 149.4  &    $ -20.1$ \\
   370     & 23:04:45.50 &  03:12:22.2 &  22.3  & 21.2 & 21.1  &  0.614 & 0.2565& 91.9   &    $ -18.4$ \\
   374     & 23:04:45.62 &  03:11:02.9 &  22.9  & 21.6 & 21.0  &  0.740 & 0.5880& 167.8  &    $ -19.9$ \\
   401$^1$ & 23:04:44.83 &  03:11:23.5 &  24.0  & 22.5 & 22.1  &  0.380 & 0.365 & 70.0   &    $- 17.9$ \\
   417     & 23:04:44.29 &  03:12:55.6 &  22.9  & 21.7 & 21.5  &  1.169 & 0.4675& 242.4  &    $ -19.2$ \\
   480     & 23:04:42.51 &  03:10:16.9 &  21.1  & 20.4 & 20.2  &  1.610 & 0.4560& 330.3  &    $ -20.5$ \\
   488     & 23:04:41.56 &  03:13:36.2 &  20.9  & 19.1 & 18.2  &  2.021 & 0.3065& 337.6  &    $ -20.9$ \\
   492     & 23:04:42.47 &  03:10:19.2 & $>$25.2& 22.0 & 20.8  &  1.579 & 0.5910& 358.6  &    $ -19.5$ \\
   509     & 23:04:41.47 &  03:08:36.3 &  22.6  & 21.0 & 20.4  &  3.283 & 0.458 & 674.8  &    $ -19.9$ \\
   548     & 23:04:40.76 &  03:09:29.3 &  23.0  & 22.0 & 21.0  &  2.512 & 0.6925& 598.8  &    $ -19.9$ \\
   593     & 23:04:38.92 &  03:11:52.8 &  22.1  & 20.8 & 20.2  &  1.514 & 0.0   &  ...   &       ...  \\
   748     & 23:04:31.99 &  03:11:18.7 &  21.6  & 20.3 & 19.4  &  3.272 & 0.2765& 513.7  &    $ -19.4$ \\
\tableline
\multicolumn{10}{l}{$^1$ This is the only object with a determined redshift within the IRCAM field of view;}\\
\multicolumn{10}{l}{we measure $J=20.8\pm0.2$ and $K=20.0\pm0.3$}
\end{tabular}
\end{center}
\end{table*}

\begin{figure}[th]
\vspace*{-2cm}\hspace*{-13mm}\psbox[xsize=0.9\textwidth]{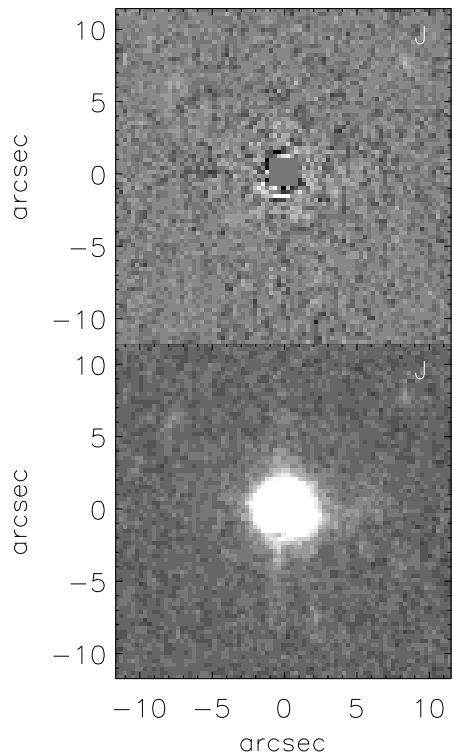}
\vspace*{-15mm}\caption{Results of our PSF subtraction in the $J$-band before
(bottom) and after (top). 
\label{fig_noprof}}
\end{figure}

An obvious question is whether there could be a galaxy aligned
precisely with the center of the QSO. We were unable to adequately
remove the QSO Point Spread Functions (PSFs) from the optical images
because the QSO profiles were saturated, and the resulting profile
subtraction was poor. Instead, we concentrated on removing the
unsaturated quasar profile from the $J$-band image, where we expect to
be most sensitive to an intervening galaxy at $z\:=\:0.7$, using the PSF of a
standard star taken immediately prior to the observations of
Q2302+029. The stellar PSF was scaled and shifted to produce a minimum
$\chi^2$ between it and the QSO profile, and the result was subtracted
from the quasar image. We set the central 7x7 pixels to zero to allow
a best fit to the profile wings alone, since this is where we are most
likely to resolve a galaxy.  The subtraction is presented in
Figure~\ref{fig_noprof}.  Some residuals are seen due to the finite
signal-to-noise of the standard star data, and do not represent the
detection of an underlying galaxy. Although the signal-to-noise of the
image is degraded from subtraction of an empirical PSF, we can still
detect the majority of the faint galaxies cataloged from the
unsubtracted frame, meaning that the magnitude limit is still
$J\:\sim\:22.6$. As Figure~\ref{fig_noprof} shows, we detect nothing 
outside a radius of $\sim 5$ pix from the center of the QSO, or 1.4
arcsec. This corresponds to 6~\h\ at $z\:=\:0.7$.

\subsection{Spectroscopy}

\begin{figure*}[ht]
\vspace*{-2cm}\centerline{\psfig{figure=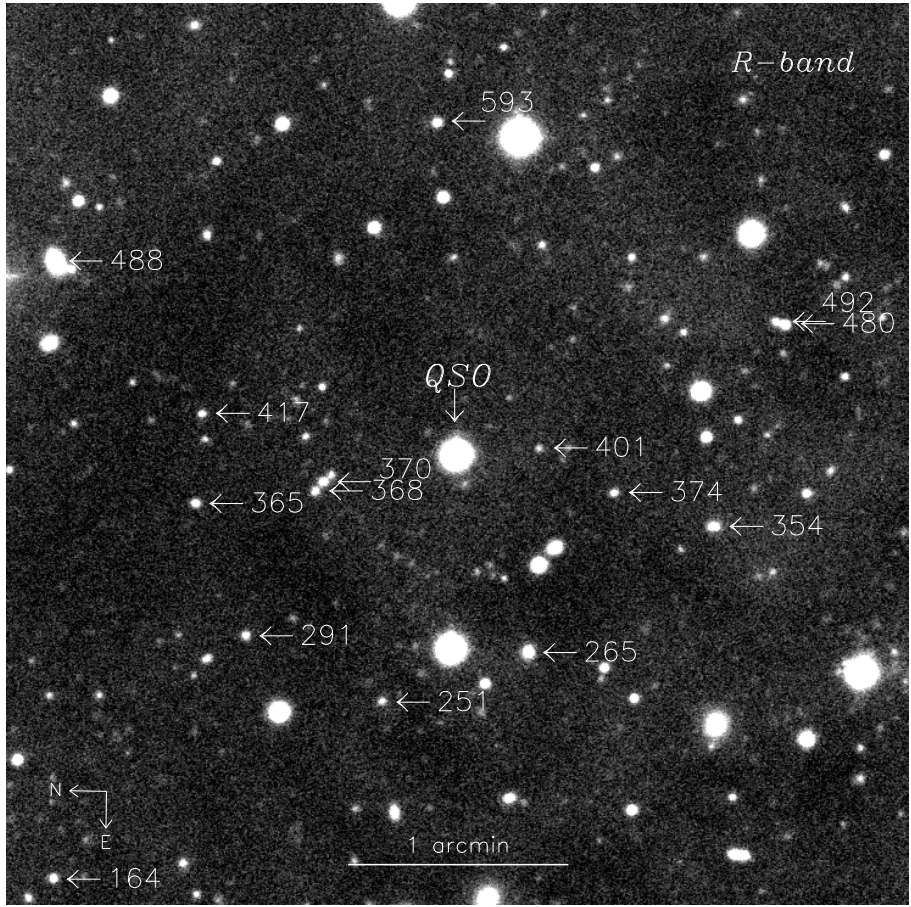,height=15cm,angle=0}}
\vspace*{-1cm}\caption{4x4 arcmin section of the $R$-band image of Q2302+029 
showing galaxies with spectroscopic
redshifts. Object designations correspond to those given in Table~1
and Table~3.\label{fig_opt}}
\end{figure*}

We were able to measure 23 `secure' redshifts in the field around
Q2302+029 with LDSS-2. These are listed in Table~3, which
gives the object ID, its position, $R$-, $I$-, and $B$-band
magnitudes, separation, $\theta$, from the QSO sightline, derived
redshift, corresponding separation, $\rho$, in \h , and resulting
absolute magnitude, $M$, without any $k$-correction applied (see \S
4.1).  The positions of these objects on the sky are shown overlaid on
the $R$-band image in Figure~\ref{fig_opt}. We note, without showing
all the galaxy spectra obtained, that these represent only the
redshifts we consider to be secure. For the majority of the galaxies
observed, we were unable to deduce a redshift, largely because we had
selected galaxies at magnitudes expected for intermediate-redshift,
but which were close to the limit of the WHT's sensitivity. The
redshifts in Table~3 are quoted with varying degrees of
precision; these crudely reflect the accuracy with which they are
measured, which depends on the quality of the data.

\begin{figure}[th]
\hspace*{0mm} \psbox[xsize=0.4\textwidth]{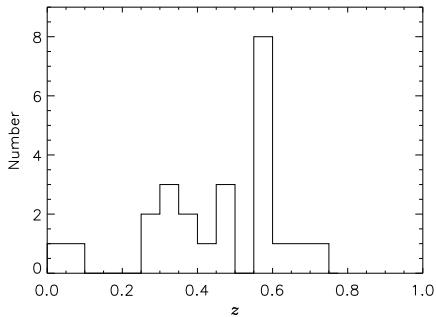}
\caption{Redshift distribution of all the galaxies observed by
LDSS-2 for which `secure' redshifts could be obtained, i.e. those
listed in Table~3. A peak at $z\:=\:0.59$ is clear.
\label{fig_dist}}
\end{figure}

We detect two galaxies at a redshift near that of the absorber,
galaxy~548 at $z\:\sim\:0.6925$, which is 599~\h\ away from the QSO
sightline, and galaxy~251 at $z\:=\:0.7197$, which is only 282~\h\
away. We discuss the likelihood that these are the absorbers in
\S 4. In Figure~\ref{fig_dist} we plot a histogram of the
redshift distribution for the 23 galaxies measured. Although the
numbers are small, we  detect an overdensity of galaxies at 
$z\:\sim\:0.59$. We also discuss this apparent cluster of galaxies in
\S 4.

We were able to obtain the redshift of one galaxy in IRCAM's small
field of view, that of galaxy~401 at $z\approx\:0.365$.
Although not at the redshift of the high ionization system,
we note in passing that the galaxy is of interest in its
own right; it lies 70~\h\ from the QSO sightline, and has a measured
absolute magnitude of $M_R\:=-\:18.2$ and $M_B\:=-\:16.8$ (after
applying a suitable $k$+EC-correction of 0.3 and 0.4 mags for $R$ and
$B$ respectively, and assuming $h\:=\:1$---see \S 4.1).  An absorption
line at 1661.6~\AA\ is identified by \citet{KP13} as ``C~III $\lambda
977$?''  at $z\:=\:0.7016$, although the difference between the
measured and expected wavelength is large, $-0.94$~\AA . We show the
expected position of \lya\ at the redshift of galaxy 401 in
Figure~\ref{fig_zoom}, and although the error on the galaxy's redshift
is large, it is possible that the absorption line at 1661.6~\AA\ is
\lya\ at $z\:=\:0.3667$.  The strength of the line and the galaxy's
luminosity are just about consistent with the relationships of
sightline separation, luminosity, column density, and equivalent width
found by \citet{Chen98} for \lya\ absorbing galaxies, even though the
line is strong (rest equivalent width $W_\lambda\:=\:1.28\pm0.12$~\AA\
for $z\:=\:0.3667$) and the galaxy faint, with a magnitude close to
that of the Small Magellanic Cloud.

                   \subsection{Photometric redshifts}

Due to the difficulty in obtaining a significant number of galaxy
redshifts at $z\:=\:0.7$ with a 4~m telescope, we developed algorithms
to constrain a galaxy's redshift photometrically.  We used the
synthetic stellar population models developed by \cite{Raul00} which
are determined by the spin, $\lambda$, of the dark halo of the galaxy,
to produce galaxy SEDs. \cite{HJ99} obtained closed analytic
expressions for the SFR and gas fraction of a galaxy and full details
can be found therein; to generalize, a value of $\lambda\:=\:0.02$
yields a SFR typical of an Sab galaxy, while $\lambda\:=\:0.10$ gives
the SFR of an irregular galaxy.

In order to compute the photometric redshift of an observed galaxy we
allowed our model parameters---age, redshift, star formation
efficiency---to take any possible value and used a $\chi ^2$
minimization to find the best fitting values. Our approach is similar
to that developed in \cite{Arno99} but is contrary to that of other
groups [e.g. \citet{CCSB99,Driv98}] who derive redshift templates from
observations and/or a mixture of observations and models.  Since the
purpose of this paper is not to derive precise redshifts for every
galaxy in the field of Q2302+029, which would be difficult considering
the small number of photometric bands used, we will not enter into a
detailed comparison of our fitting procedure with other redshifts
surveys. We note, however, that we have made such comparisons, and
that we were able to recover measured redshifts (such as those in the
Hubble Deep Field) to an accuracy in redshift of $\Delta z \sim 0.1$.

As described in \S 1, these algorithms work best when a galaxy is
observed in the infra-red, when the flux is observed at wavelengths
beyond the peak of its SED.  Below, therefore, we discuss the five
brightest galaxies seen close to the QSO (that is, within the IRCAM
field of view) that have no measured redshift and discuss the
possibility that they could be at a $z\:\sim\:0.7$.  First we note a
general principle.  In Figure~\ref{fig_sed1} we plot the SED for a
typical galaxy at $z\:=\:0$ and the same (younger) galaxy at
$z\:=\:0.7$. We also overplot the relative flux limits obtained from
our images (diamonds) arbitrarily normalized such that
$F_\lambda$($I$) is close to the peak of the SED of the galaxy at
$z\:=\:0.7$. The figure shows that its extremely difficult for any
galaxy we detect in all four bands to be at a redshift of $0.7$,
essentially because the blue flux from the galaxy is predicted to be
low. Only if the galaxy has a particularly high SFR can the blue flux
be detected in the $B$-band from a galaxy at intermediate redshift.

\begin{figure}[th]
\hspace*{0mm}\psbox[xsize=0.45\textwidth]{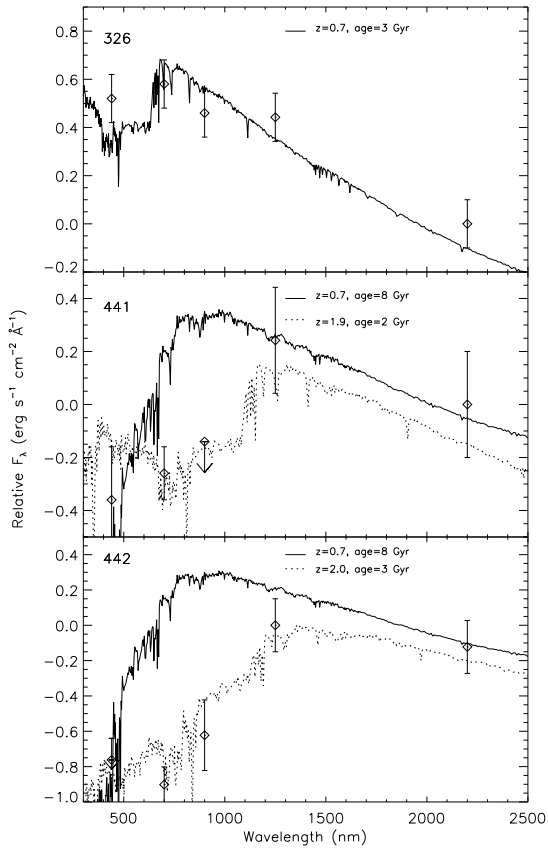}
\caption{Comparison of theoretical SEDs to the photometry of the
four brightest galaxies in IRCAM field of view. In all cases, we have
constrained the SED to be at a redshift of 0.7 and varied the galaxy's
age, metallicity and SFR to best fit the data
points (solid line). 
\label{fig_sedfits}} 
\end{figure}

{\it 326} --- The derived photometric redshift is
$z_{\rm{phot}}\:\sim\:0.05$.  The optical images suggest that the
galaxy is elliptical, while in the IRCAM images it appears more
irregular. Archival HST WFPC2 images of the galaxy clearly reveal some
spiral structure, although the overall morphology appears
irregular. The galaxy is detected in all bands;
Figure~\ref{fig_sedfits} shows that the photometric redshift is ill
defined since the continuum is quite flat, and suggests that it might
be possible for the galaxy to be at $z\:=\:0.7$ if very
young. However, the unambiguous resolution of structure in the WFPC2
image clearly demonstrates that the galaxy must be at a fairly low
redshift and cannot be at $z\:\sim\:0.7$. The galaxy would also be
unusually bright: applying suitable $k$- and EC-corrections (see \S
4.1) would give absolute magnitudes of $M_R\:=\:-22.5$ and
$M_B\:=\:-21.3$.

{\it 320} --- {\tt sextractor} defines the object to be a star in both optical
and IR images. This is confirmed in the WFPC2 image.

{\it 441 \& 442} --- As shown in Figure~\ref{fig_sedfits} the relative
distribution of the photometric points for both galaxies are
completely incompatible with any galaxy at $z\:\sim\:0.7$.  Both must
lie at $z> 1$, and to further demonstrate this, we include best-fit
SEDs to the data points: for galaxy 441, $z_{\rm{phot}}\:\sim\:1.9$,
while for 442 $z_{\rm{phot}}\:\sim\:2.0$.

{\it 382} --- This galaxy is just detected in $R$ and so is included
in our catalog, but it is not detected by {\tt sextractor} in the
remaining bands. In part, this is due to a combination of its
faintness combined with its proximity to the QSO, where the background
is higher, and in fact a visual inspection of the frames suggest that
the galaxy might just be detected in $B$, $J$ \& $K$.  Although we
cannot calculate a photometric redshift from only one color, we think
it unlikely that the object could be at $z\:=\:0.7$: the galaxy would
have to be very young ($\leq 2$ Gyr old), bright, with
$M_R\:=\:-19.9$, and have an extremely high star formation rate to
explain the detection in the $B$-band. Although these constraints are
not impossible to satisfy in principle, they would be unusual, and
with no reliable photometry, we can make no claim as to the galaxy
being responsible for the high ionization system.

                       \section{Discussion}

       \subsection{The $z\:=\:0.7$ high-ionization absorption system}

We have been unable to detect any obvious bright galaxies within 30
arcsec of Q2302+029 which could be responsible for the $z\:=\:0.7$
high-ionization absorption system detected by Jannuzi~\etal\ (1996).
Our magnitude limits are given in Table~2; to
estimate the absolute magnitudes that these observed limits correspond
to, we have calculated the cosmological dimming, $M_{\rm{obs}}$, then
applied both a $k$- and Evolutionary-correction (EC) to derive the
absolute limiting magnitude, $M$, in individual bands.  Besides having to
correct a galaxy's magnitude in a particular band for its redshifted
SED (the standard $k$-correction), the galaxies observed at
$z\:\sim\:0.7$ are younger than they would be now. So to compare
galaxy magnitudes with those in the nearby universe, an EC-correction
is also required.  We have used our SED models discussed in \S 3.3 to
estimate a combined $k$- and EC-correction; we have taken three values
for the spin parameter of the halo ($\lambda=0.01,0.05$ and 0.1) which
represent all possible star formation histories in our models---from
ellipticals to irregular types of galaxies. We have then formed each of the
types at $z > 2$ and computed the difference for each band between its
$z=0$ value and the redshift of interest, resulting in the desired
$k$+EC-correction. The correction obviously depends on the redshift of
formation of the galaxy and its SFR. We find that for the IR bands the
SFR and formation redshift are irrelevant since they sample mostly the
old population, while in the optical the $B$-band $k$+EC corrections
vary the most for different galaxy types: the maximum
spread in the correction is $\sim\:0.5$ mags in the $B$-band, compared
to only $\sim\:0.1$ in the $K$-band.

Table~2 also lists present day values of $M^*$
[corresponding to the $L^*$ break in the luminosity function
\citep{Shec76}] for each band, taken from various galaxy surveys. We
have been unable to find a value of $M^*(J)$ in the literature, so we
have used the value of $M^*(K)$ given by \citet{Love00} and assumed
$J-K\:=\:0.8$. The final column gives the resulting value of
$L/L^*$. In the optical we are able to find galaxies down to
luminosities of $\sim\:1/10 L^*$, while in the $J$-band we are
sensitive to galaxies with  $L\:\sim\:1/20 L^*$ or greater.

In our spectroscopic survey around the QSO, we detect two galaxies at
a redshift near that of the absorber, galaxy~548 at $z\:\sim\:0.6925$,
which is 599~\h\ away from the QSO line of sight, and galaxy~251 at
$z\:=\:0.7197$ which is only 282~\h\ away.  Whether the detection of
these two galaxies in our survey marks the existence of a cluster at
$z\:=\:0.7$ is unclear. At this redshift, the sensitivity of
our survey (as a function of magnitude limit and volume sampled) is
falling rapidly, and the detection of two galaxies at $z\:=\:0.7$
might correspond to an `over-density' of galaxies. Unfortunately, our
survey was not designed to be complete to any specified magnitude
limit, and we cannot calculate the sensitivity function of the survey
to estimate whether the detection of two galaxies at the limit of
detectability is significant.

The difference in redshift between galaxy 251 and the narrow O~VI
system at $z\:=\:0.7106$ is $\simeq\:1600$~\kms . The error in the
redshift measurement is of order 100$-$200~\kms, so the velocity
difference between galaxy and absorber is large, probably ruling out
the galaxy as being associated with the absorber.  The separation is
also large compared to those found for \lya -absorbing galaxies. If
the O~VI cross-section of a galaxy was larger than that found by
\cite{Chen98} for a \lya\ halo, i.e., $>\:160$~\h , then the
ionization at the edge of the halo would have to be sufficiently
extreme to ionize the H~I and produce O~VI. In principle, the
background UV radiation field can induce O~VI ions in sufficient
number to be produce the types of column densities detected, but the
gas density must be low (i.e., the ionization parameter---the ratio of
ionizing photons to total ionized+non-ionized hydrogen atoms---must be
high) and/or the metallicity of the gas must be high \citep[see, e.g.,
the discussion by][]{trip00b}. This means that the path length must be
extremely long, of order several hundred kpc. To obtain such a path
length at the edge of a halo would require the radius to be $>>200$~\h
, at which point the distinction between halo and the general IGM is
severely blurred.  Hence it seems unlikely that galaxy 251 is directly
responsible for the absorption.

The remaining mechanisms which might be responsible for the
high-ionization system towards Q2302+029 have been discussed by
\citet{Janu96}. To summarize, the absorption might arise from gas
associated with: a) supernova ejecta in a galaxy along the line of
sight, b) high velocity material ejected from the QSO, or c)
intracluster material.  Our observations suggest that there is
unlikely to be a galaxy close to the line of sight in which a
supernova occurred, although in principle the host could be perfectly
aligned with the QSO image and not detected even with our subtraction
of the QSO PSF in the $J$-band. Since the strength and breadth of
absorption from supernovae ejecta decreases with time, this
explanation can be easily rejected by re-observing the high ionization
system and confirming that the system still exists.

The lack of any bright galaxy close to the sightline actually supports
both remaining hypotheses.  In general, narrow absorption line systems
have been considered unrelated to the QSO and its environs if $\Delta
v\: >\: 5000$~\kms ; even though the existence of an absorption system
$\sim\:56,000$~\kms\ from the emission redshift of the QSO challenges
our conventional view of the origin of metal-line absorption systems,
the absence of any galaxies close to the line of sight is an obvious
corollary. There is also some precedent for such an explanation in
recent work by \citet{Hama97}, which details the variability of both
broad and narrow C~IV and Si~IV lines towards another QSO
Q2343+125. In this case, the velocity difference between QSO and
absorbing gas is $\simeq\:24,000$~\kms, and the variability is the
strongest evidence that the system is actually intrinsic to the QSO.
Clearly, searching for variability in the absorption lines towards
Q2302+029 will be a powerful test of whether the $z\:=\:0.7$ system
is intrinsic. Such data may soon be available (Januzzi~\etal , in
preparation).

The third possibility suggested above, that we are seeing absorption
by intracluster gas, is also problematic. If the absorption is due to
the warm/hot gas which occurs in shocked regions collapsing with the
dark matter, as suggested by the models, then the width of the
absorption represents a cosmological distance, rather than an intrinsic
velocity dispersion from thermal or turbulent motions.  Thus, if the
width of the structure is $\sim\:4000$~\kms\ wide, then 
the corresponding cosmological distance would be $\sim\:40$~\mpc\ (comoving).
This would imply a rather special geometry, with 
the sightline  passing {\it along} a
filament---a rare, but not impossible alignment.  If true though,
would we not expect to see a high surface density of galaxies, due to the
alignment?  The hydrodynamical simulations which exist at present do
not have sufficient resolution to estimate the number of galaxies
along a shock front, but we can crudely estimate the surface density
of galaxies we would expect to see in our IRCAM field of view. If we
assume that the luminosity function (LF), $\phi$, of galaxies at
$z\:=\:0.7$ is the same as the Schechter LF of the local universe, then
we can simply integrate the LF

\begin{equation}
d\phi\:=\:0.92\:\phi^*\:(L/L^*)^{\alpha+1} \exp(-L/L^*)\: dM
\end{equation}

between the observed magnitude limits and arbitrarily bright
magnitudes. Here, $\phi^*$ is the number of galaxies with luminosity
$L^*$ and $\alpha$ parameterizes the shape of the function at
magnitudes fainter than $L^*$. For the $K$-band, \citet{Love00}
derives values of $\alpha\:=\:-1.16$, $M^*$($K$)$\:=\:-23.6$, and
$\phi^*\:=\:0.012\: h^3$~Mpc$^{-3}$. Integrating equation 1 using a
minimum magnitude of $M$($K$)$\:=-21.0$ gives a total of
$\sim\:0.03\:h^3$~Mpc$^{-3}$. Similar values are derived for the
$J$-band. This would imply that the surface density of galaxies in the
40~\mpc\ filament would be $\sim\:1$ $h^{2}$ galaxy Mpc$^{-2}$. Yet
the field of view studied with IRCAM is only $\sim\:0.06
h^{-2}$~Mpc$^{2}$---hence we clearly expect $<<\:1$ galaxy over such a
small field of view, consistent with our observations. Unsurprisingly,
our pencil-beam along the sightline is simply too narrow to detect
significant over-densities of galaxies.  It thus remains possible that
the sightline to Q2302+029 passes through a warm/hot gas filament
along its length.

\begin{figure*}[ht]
\hspace*{10mm}\psfig{figure=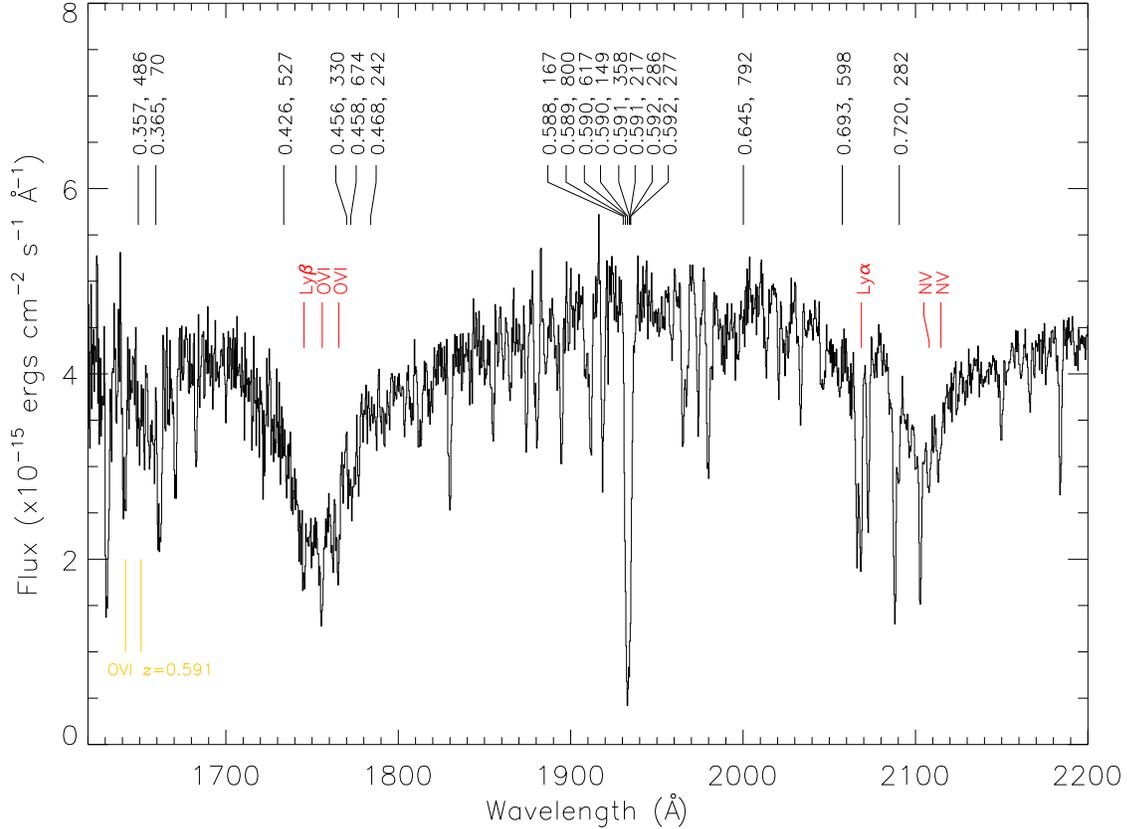,height=12cm,angle=90}
\caption{FOS spectrum of Q2302+029 \citep[see][]{Janu96}. The
positions of the narrow \lya, Ly$\beta$, O~VI and N~V lines at
$z\:=\:0.7106$ are shown just above the spectrum, superimposed on the
broader lines discussed in the text. Above these, the wavelengths
predicted for \lya\ from the galaxies in Table~3 are plotted, along with
their redshift and their separation from the line of sight in \h . The
\lya\ absorption at the cluster redshift is clear. Below the spectrum,
we also plot the expected position of O~VI from the cluster at $z\:=\:0.59$. This
is shown more fully in Figure~\ref{fig_zoom}
\label{fig_spec1}}
\end{figure*}

Finally, with regard to our search for O~VI systems in general, we
cannot say whether our results towards Q2302+029 are indicative of the
population of the O~VI system as a whole. The absorption towards
Q2302+029 is clearly pathological, since none of the other
(apparently) intervening $z\:<\:1$ O~VI systems detected so far are
accompanied by the broad, high-ionization absorption systems seen
towards Q2302+029. Only further imaging and spectroscopy of fields
around QSOs showing $z\:<\:1$ O~VI systems will reveal whether
individual galaxies are responsible for O~VI systems, as well as
searches for O~VI absorption in the spectra of background QSOs located
near foreground galaxies.

         \subsection{The cluster at $z\:\sim\:0.59$ and associated
                               absorption lines}

Although we detect few galaxies close to the QSO sightline, our
spectroscopic survey shows that we have detected a cluster of galaxies
at $z\:\sim\:0.59$.  In Figure~\ref{fig_spec1} we reproduce a portion
of the G190H FOS spectrum published by \cite{Janu96} and plot the
positions of \lya\ absorption expected at the redshift of each galaxy,
along with its separation from the QSO line of sight (in \h ).  \lya\
is detected at $z\:=\:0.5904$, and is strong: \citet{KP13} measure a
rest equivalent width of 1.25~\AA , but suggest that there may be
contamination by Ly$\delta$ from a system at $z\:=\:1.0361$. This
higher-$z$ system is detected in \lya, Ly$\beta$, Ly$\gamma$ and
Ly$\epsilon$, making it relatively easy to calculate the expected
equivalent width of Ly$\delta$. A simple curve of growth analysis is
well able to reproduce the Lyman equivalent widths (to within the
quoted errors) with $\log$($N$)$\:\simeq\:15.6$ and
$b\:\simeq\:55$~\kms, meaning that we expect
$W_\lambda$(Ly$\delta$)$\:\simeq\:0.3$~\AA . Simply subtracting this
from the total equivalent width of the $z\:=\:0.5904$ \lya\ line
leaves $W_\lambda$(\lya )$\:\sim\:1.0$~\AA .  This is a relatively
strong \lya\ line, and if the correlation of \lya\ equivalent width
and galaxy separation found by \cite{Chen98} for \lya -absorbing
galaxies were to hold here, we would expect to find at least one
galaxy at a separation of $\approx 30-100$~\h , and $L/L^*\:> 0.3$,
i.e. well within our detection limits. We tentatively detect one
galaxy at a separation of 149~\h\ at $z\:=\:0.59$
(Table~3) which could in principle be responsible for the
absorption. However, given the galaxy's separation, the line probably
has too large an equivalent width to well fit the correlation found by
\cite{Chen98}.

\citet{orti99} provide an example of \lya\ absorption associated with
a group of galaxies towards Q1545+2101, where absorption may occur
not from intragroup gas but from the overlapping halos of individual
galaxies comprising that group. Towards Q2302+029, it would seem that
we have an example where strong absorption is clearly associated with
a cluster of galaxies, but {\it not} with the halos of individual galaxies
within that cluster. The comparison with Q1545+2101 is particularly
interesting.  The galaxies associated with the absorption towards
Q1545+2101 range in absolute magnitude from $M_R - 5 log h\:= -19.0$
to $-20.8$. As shown in Table~2, we would easily be able
to detect these galaxies in $R$ (and in the IR), so our inability to
detect absorbing galaxies towards Q2302+029 is not because of
insufficient sensitivity.\footnote{A difference in redshift of 0.1,
from $z\:=\:0.7$ in Table~2, to $z\:=\:0.59$ for the
cluster, results in an {\it increase} in sensitivity by 0.4 mags} The
most important piece of evidence linking the \lya\ absorption towards
Q1545+2101 and the galaxies near the QSO sightline is the
multicomponent structure of the \lya\ line. \citeauthor{orti99} found eight
individual \lya\ components spanning a total range of 535~\kms .  It
would be interesting to see if the \lya\ line arising at the same
redshift as the cluster towards Q2302+029 also shows multicomponent
structure at high resolution; clearly, smoother intragroup gas might
not have the same complex structure as that seen towards Q1545+2101.
We note, however, that the system towards Q1545+2101 may not be a good
example of absorption from a group of intervening galaxies, since the
redshifts of the galaxies are the same as the QSO, i.e. this is an
intrinisic system. The origin of the absorption systems in these
systems is far from established, and the observed \lya\ line may have
nothing to do with galaxy halos.

\begin{figure}[th]
\hspace*{0mm} \psbox[xsize=0.5\textwidth,rotate=l]{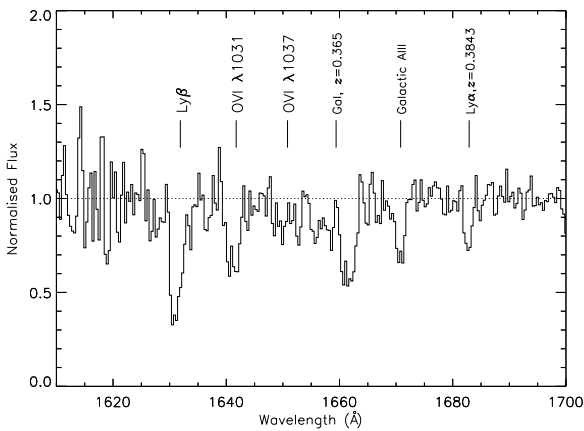}
\caption{Detail of the FOS spectrum shown in the previous figure,
centered at the expected position of O~VI at the cluster's redshift
of $z\:=\:0.59$. In this case, the spectrum has been
normalized. Ly$\beta$ is clearly detected, and a line probably
corresponding to the O~VI $\lambda 1031$
line is clearly visible. We also label the expected wavelength of \lya\
from the one galaxy in the IRCAM field of view whose redshift  we were
able to measure. This galaxy, 401, at $z\:\approx\:0.365$ is 70~\h\ from
the line of sight, and is discussed in Section 3.2
\label{fig_zoom}}
\end{figure}

The absorbing gas at $z\:=\:0.59$ is also a high-ionization system.
In Figure~\ref{fig_zoom} we reproduce a detail of the FOS spectrum
around the area where O~VI and Ly$\beta$ are expected, and show that
the detections appear secure. There is unfortunately no corresponding
O~VI $\lambda 1037$ line to confirm the identification, although this
is plausibly lost in the noise. We note that there is no detection of
N~V at this redshift, even though the continuum is sufficiently clear
to show any moderately strong lines, but \citet{KP13} do list the
detection of Si~III $\lambda 1206$, Si~IV $\lambda 1393$, and both
C~IV lines. Hence it appears that we have a second example of O~VI
absorption without any obvious nearby galaxies, at least within the
$\sim160$~\h\ radius found by \cite{Chen98} for \lya -absorbing
galaxies, unless the galaxy 149~\h\ away contributes to the
absorption. Only a more complete search towards other sightlines will
reveal whether high-ionization systems are associated with galaxies at
such large distances. 

Again, it is possible that a cluster galaxy sits exactly on top of the
QSO, and cannot be seen in our subtraction of the QSO profile. If so,
we would now require that there exist {\it two} galaxies hidden from
view (including the $z\:\sim\:0.7$ O~VI absorber). This may be
stretching the plausibility of this caveat.  Absorption could also
occur in a cluster galaxy fainter than our detection limits---this is
a problem which has plagued the search for absorbing galaxies at all
redshifts beyond $z\:=\:0$, and our analysis is no different in being
unable to address the problem beyond establishing secure magnitude
limits in multiple photometric bands. Obviously, deeper images and
more detailed spectroscopy should be possible with larger aperture
telescopes.  If we really are seeing intracluster gas not associated
with any individual galaxy at $z\:=\:0.59$, then higher resolution
observations, allowing calculation of accurate column densities, may
be useful in determining the physical conditions in the gas, which can
in turn be compared with the predictions from the numerical simulations of
the evolution of the warm/hot gas component.

\acknowledgments

We'd like to thank the many support astronomers who have helped us at
all the telescopes used to obtain the data, and the WHT TAC who
awarded us a substantial allocation of LDSS-2 time. Thanks also to
Todd Tripp for helpful discussions. Support for this work was provided
in part through grants GO-06707.01 and GO-08165.01 from the Space
Telescope Science Institute, which is operated by the Association of
Universities for Research in Astronomy, Inc., under NASA contract
NAS5-26555.

\clearpage

\bibliographystyle{apj}
\bibliography{apj-jour,bib1}

\clearpage

\clearpage

\begin{table}
\caption{Photometry of galaxies detected near the sightline of Q2302+029 \label{tab_close}}
\begin{center}
\begin{tabular}{cccrrrrrc}
\hline\hline
    &\multicolumn{2}{c}{Position (J2000.0)} &&&&&&$\theta$ \\
ID    & RA  & DEC &  \push{$B$} & \push{$R$} & \push{$I$} & \push{$J$} & \push{$K$} & (arcmins) \\
\hline
   320 & 23:04:46.96 &  03:11:23.1 & 20.9     & 18.4     & 17.0     & 16.0    & 15.1      &  0.628 \\
   326 & 23:04:46.64 &  03:11:18.8 & 21.6     & 20.0     & 19.5     & 18.5    & 17.4      &  0.618 \\
   330 & 23:04:47.21 &  03:11:32.5 & $>$25.2  & 22.5     & 20.9     & ...     & ...       &  0.604 \\
   338 & 23:04:47.16 &  03:11:35.5 & $>$25.2  & 23.1     & $>$22.8  & ...     & ...       &  0.575 \\
   342 & 23:04:47.10 &  03:11:40.0 & $>$25.2  & 22.5     & 22.9     & ...     & ...       &  0.542 \\
   345 & 23:04:46.98 &  03:11:36.5 & $>$25.2  & 22.6     & 22.4     & ...     & ...       &  0.527 \\
   347 & 23:04:47.19 &  03:11:53.3 & $>$25.2  & 24.0     & $>$22.8  & ...     & ...       &  0.567 \\
   349 & 23:04:46.96 &  03:11:57.8 & 24.5     & 23.7     & $>$22.8  & ...     & ...       &  0.533 \\
   350 & 23:04:47.12 &  03:11:50.9 & 25.7     & 24.6     & $>$22.8  & ...     & ...       &  0.542 \\
   351 & 23:04:46.83 &  03:12:01.8 & 24.2     & 23.5     & $>$22.8  & $>$22.6 & $>$21.1   &  0.532 \\
   357 & 23:04:44.97 &  03:11:46.2 & 15.8     & 15.5     & 15.2     & 14.7    & 13.8      &  0.000 \\
   358 & 23:04:46.65 &  03:12:06.3 & 24.0     & 23.9     & $>$22.8  & 22.1    & 19.4      &  0.537 \\
   364 & 23:04:46.36 &  03:11:55.1 & $>$25.2  & 24.3     & $>$22.8  & $>$22.6 & $>$21.1   &  0.377 \\
   370 & 23:04:45.50 &  03:12:22.2 & 22.3     & 21.2     & 21.1     & ...     & ...       &  0.614 \\
   373 & 23:04:46.00 &  03:11:14.8 & $>$25.2  & 24.0     & $>$22.8  & ...     & ...       &  0.583 \\
   382 & 23:04:45.51 &  03:11:43.6 & $>$25.2  & 22.6     & $>$22.8  & $>$22.6 & $>$21.1   &  0.142 \\
   385 & 23:04:45.38 &  03:12:20.1 & $>$25.2  & 22.1     & 20.9     & ...     & ...       &  0.574 \\
   396 & 23:04:45.03 &  03:11:17.0 & 24.4     & 23.4     & $>$22.8  & ...     & ...       &  0.487 \\
   401 & 23:04:44.83 &  03:11:23.5 & 24.0     & 22.5     & 22.1     & 20.8    & 20.0      &  0.380 \\
   403 & 23:04:44.82 &  03:11:15.5 & 24.4     & 22.8     & 21.5     & ...     & ...       &  0.513 \\
   411 & 23:04:44.57 &  03:11:53.9 & $>$25.2  & 23.4     & $>$22.8  & $>$22.6 & $>$21.1   &  0.163 \\
   419 & 23:04:44.40 &  03:12:21.8 & 24.4     & 23.4     & 22.3     & ...     & ...       &  0.610 \\
   423 & 23:04:44.64 &  03:12:12.5 & $>$25.2  & 24.5     & $>$22.8  & $>$22.6 & $>$21.1   &  0.446 \\
   428 & 23:04:44.44 &  03:11:37.2 & 24.9     & 23.9     & $>$22.8  & 22.6    & $>$21.1   &  0.200 \\
   439 & 23:04:43.92 &  03:11:31.9 & $>$25.2  & 23.1     & $>$22.8  & 22.6    & 20.9      &  0.354 \\
   441 & 23:04:43.63 &  03:11:53.8 & 25.3     & 23.6     & $>$22.8  & 20.5    & 18.9      &  0.358 \\
   442 & 23:04:43.93 &  03:11:44.2 & $>$25.2  & 24.1     & 22.6     & 20.0    & 18.1      &  0.262 \\
   484 & 23:04:42.99 &  03:11:20.6 & 24.3     & 24.0     & $>$22.8  & ...     & ...       &  0.653 \\
   493 & 23:04:42.80 &  03:11:40.2 & $>$25.2  & 24.4     & $>$22.8  & ...     & ...       &  0.551 \\
   494 & 23:04:42.72 &  03:11:42.6 & 24.8     & 24.2     & $>$22.8  & ...     & ...       &  0.565 \\
   925 & 23:04:43.49 &  03:11:36.4 & 25.9     & $>$24.5  & $>$22.8  & $>$22.6 & $>$21.1   &  0.404 \\
   932 & 23:04:42.50 &  03:11:32.7 & 25.1     & $>$24.5  & $>$22.8  & ...     & ...       &  0.656 \\
   992 & 23:04:44.83 &  03:11:51.0 & $>$25.2  & $>$24.5  & $>$22.8  & $>$22.6 & 19.7      &  0.087 \\
   993 & 23:04:46.02 &  03:11:42.6 & $>$25.2  & $>$24.5  & $>$22.8  & 22.0    & 19.8      &  0.269 \\
   994 & 23:04:45.94 &  03:11:46.2 & $>$25.2  & $>$24.5  & $>$22.8  & 22.2    & 20.8      &  0.242 \\
   995 & 23:04:44.38 &  03:11:33.6 & $>$25.2  & $>$24.5  & $>$22.8  & 21.8    & 20.9      &  0.256 \\
   996 & 23:04:43.60 &  03:11:28.4 & $>$25.2  & $>$24.5  & $>$22.8  & 21.0    & 19.0      &  0.453 \\
   997 & 23:04:43.41 &  03:11:24.2 & $>$25.2  & $>$24.5  & $>$22.8  & $>$22.6 & 19.2      &  0.535 \\
   998 & 23:04:46.28 &  03:11:20.3 & $>$25.2  & $>$24.5  & $>$22.8  & $>$22.6 & 19.8      &  0.542 \\
   999 & 23:04:42.82 &  03:11:20.7 & $>$25.2  & $>$24.5  & $>$22.8  & ...     & ...       &  0.685 \\
  1000 & 23:04:45.82 &  03:12:22.4 & $>$25.2  & $>$24.5  & $>$22.8  & ...     & ...       &  0.640 \\
\hline
\end{tabular}
\end{center}
\end{table}

\end{document}